\documentclass[epj]{svjour}
\usepackage{graphicx,graphics}
\usepackage{amsmath,amssymb,latexsym}
\usepackage{amsfonts}
\usepackage{color}
\usepackage{epsfig}
\usepackage{mathrsfs}
\usepackage{changes}
\usepackage{tikz}

\usetikzlibrary{patterns}
\usetikzlibrary{decorations}
\usetikzlibrary{%
  lindenmayersystems,
  decorations.pathmorphing,
  decorations.markings,
  shapes.geometric,
  calc%
}
\newcommand{\id}{\textrm{d}}

\def\beq{\begin{equation}}
\def\eeq{\end{equation}}
\def\bea{\begin{eqnarray}}
\def\eea{\end{eqnarray}}
\def\ba{\begin{array}}
\def\ea{\end{array}}
\def\n{\nonumber}
\def\c{\mathscr}
\def\la{\langle}
\def\ra{\rangle}

\def \D{\Delta}

\begin{document}
\title{Thermal response in driven diffusive systems}
\author{Marco Baiesi\inst{1,2} \and Urna Basu\inst{3} \and Christian Maes\inst{3}}     
\institute{Department of Physics and Astronomy, University of Padua, Via Marzolo 8, I-35131 Padova, Italy
 \and INFN, Sezione di Padova, Via Marzolo 8, I-35131 Padova, Italy
\and Instituut voor Theoretische Fysica, KU Leuven, Belgium}
%
%
\abstract{
Evaluating the linear response of a driven system to a change in environment temperature(s) is essential for understanding thermal properties of nonequilibrium systems.  The system is kept in weak contact with possibly different fast relaxing mechanical, chemical or thermal equilibrium reservoirs.  Modifying one of the temperatures creates both entropy fluxes and changes in dynamical activity.  That is not unlike mechanical response of nonequilibrium systems but the extra difficulty for perturbation theory via path-integration is that for a Langevin dynamics temperature also affects the noise amplitude and not only the drift part. Using a discrete-time mesh adapted to the numerical integration  one avoids that ultraviolet problem and we arrive at a fluctuation expression for its thermal susceptibility.  The algorithm appears stable under taking even finer resolution.
\PACS{
      {05.70.Ln}{Nonequilibrium and irreversible thermodynamics}   \and
      {05.20.-y}{Classical statistical mechanics}\and
      {05.10.Gg}{Stochastic analysis methods}\and
      {05.40.Jc}{Brownian motion}
     } 
} 
\maketitle

\section{Introduction}
A system can be studied for mechanical, chemical or thermal response depending on the stimulus or the type of reservoirs to which the system is opened.  The standard (equilibrium) fluctuation--dissipation theorem  equally relates all these responses to the equilibrium correlation  between the observable in question and the entropy flux created by the perturbation.  In particular, the change in energy of a thermally open system to a change of temperature (fixed volume heat capacity) is directly related to the system energy fluctuations or to the variance of the entropy change.

 The question of thermal response is also meaningful  for open systems in contact with different reservoirs, some of which are equilibrium heat baths with their own fixed temperature, or for Brownian particles subject to non-conservative forces while kept in a thermal environment. We then have driven systems, where one would still like to express the thermal susceptibility (to a change of one reservoir temperature) in terms of unperturbed correlation functions between observables of the system's trajectory.   It is thus part of the general ambition of formulating linear response in nonequilibrium systems, as was intensively  studied recently; see \cite{bai13} for a review. An application of such an approach is to study the dependence on reservoir temperature of heat, as described via heat capacities and thermal conductivities \cite{bok11,pes,mand,hatcap,Dhar,Lepri,rus}.  

A difficulty arising in diffusive systems, which so far eluded further statistical studies of nonequilibrium calorimetry for mesoscopic systems, is that  temperature also specifies noise amplitudes and, therefore, changing the noise makes the perturbed and the original process very incomparable.  The reason is already plain from inspecting two Brownian motions with different diffusion constants: the temporal-spatial scales of variation are quite distinct in the long run, which mathematically amounts to saying that their processes are not absolutely continuous with respect to each other. That singularity is a problem for perturbation theory, especially when using the path--integration formalism, where one needs to make sense of a density on path--space 
relating the perturbed with the unperturbed dynamics.

The present paper aims at solving by an appropriate `regularization' the problem of thermal response in nonequilibrium diffusive systems described by Langevin equations.  The point is that the singular nature of white noise is self-inflicted as an idealization or limit of reservoir properties.  The challenge is then to remain away from the delta-correlations in the white noise, and to introduce a temporal ultraviolet cut-off $ N^{-1}$ (using an analogy with field theory) which is compatible with the numerical or observable resolution.  In the response will indeed appear the rescaled correlation function $N\,\langle O; {\cal T}_N\rangle$ between the observable $O$ and the quadratic variation ${\cal T}_N = \sum_i (B(t_{i+1})- B(t_i))^2$ (sum over temporal grid with mesh $N^{-1}$) of the Brownian path $B(s)$ over $[0,t]$, rescaled with the inverse $N$ of the cut-off time. The quadratic variation ${\cal T}_N$  as such converges to $t$ in probability, but as the cut-off $N\uparrow \infty$ is removed the rescaled 
quadratic variation $N\,({\cal T}_N - t)\sim \sqrt{N}$ fluctuates wildly.  However in the correlation function $N\,\langle O; {\cal T}_N\rangle$, the rescaled quadratic variation enters locally (in time): as we have checked numerically, that procedure is stable when adding more information or measurement points to the observable.  In other words, the result does not depend on the coarse-graining when sufficiently fine and there appears a well-defined limit of vanishing cut-off, which however we do not control mathematically. 
Nevertheless the limit makes sense if only the observable function itself is also consistently described according to the chosen path-discretization, keeping in mind that the discretization itself may very well depend on the temperature that one is perturbing.
The result is an 
expression for the thermal response in terms of a correlation function between observable and a typical nonequilibrium expression where both excesses in entropy flux and in dynamical activity play the leading role.

The technical aspects of this work are particularly useful for evaluating thermal response in diffusive systems via numerical integration, which is important to start statistical mechanical discussions of nonequilibrium calorimetry.  We concentrate on the set-up of Markov diffusion processes, first as models for mesoscopic particle motion (weakly dependent driven colloids) and secondly as models for heat conduction, e.g.~using oscillator chains.

The plan of the paper is as follows. The next section  explains the problem of nonequilibrium thermal response from a more general perspective.  In Section \ref{sec:result} we illustrate our result with the example of a boundary driven Fermi-Pasta-Ulam chain.
A detailed derivation of our new results and thermal response formul{\ae} in terms of fluctuations are found in Section \ref{sec:thre}.  

\section{The problem}\label{sec:problem}
Linear response opens a wealth of opportunities for characterizing the nonequilibrium condition but its physical interpretation is not straightforward. 
Various ways have been suggested for systematic unification also addressing the general physical meaning and usefulness~\cite{bai13,lip05,Speck,mar08}. Indeed, as we are formally dealing with a seemingly simple first order perturbation theory, attention shifts to what are the physically most reasonable choices from a plethora of correct response expressions.

\subsection{The problem with the Agarwal--Kubo approach for nonequilibrium purposes}
It is instructive to illustrate part of a first problem for nonequilibrium response with a well-known formulation by Agarwal in 1972 following Kubo's derivation for equilibrium, and rediscovered later in similar forms \cite{ag,kub,bai13}. Let us consider a Markov process  with probability density $\rho_s$ at time $s\leq 0$ satisfying the Fokker-Planck equation as summarized via the forward generator $L^\dagger$,
\[
\frac{\id}{\id s} \rho_s = L^\dagger \rho_s,\quad L^\dagger\rho =0
\]
$\rho$ being a smooth stationary density. The process gets perturbed at time zero and that generator $L^{\dagger}$ changes into 
\bea
L^{\dagger}_\varepsilon  \equiv L^{\dagger} + \varepsilon L^\dagger_{\rm pert} \label{eq:pert}
\eea
where $\varepsilon$ is a small parameter dictating the amplitude of the perturbation per unit time.
 The perturbation is switched on at time $t=0$ having an effect such as for system observable $O$ whose expectation moves from $\langle O(0)\rangle_0$ at time zero to $\langle O(t)\rangle_\varepsilon$ at time  $t>0$.  The formal result of a first order Dyson expansion is
 \begin{equation}
\langle O(t)\rangle_\varepsilon - \langle O(0)\rangle_0 = \varepsilon \int_0^t\, \left\langle \frac {L_{\rm pert}^\dagger\rho}{\rho}(0) \,O(s)\right\rangle_0 \, \id s\label{eq:ak}
\end{equation}
in terms of a time-correlation function for the unperturbed process.
This Agarwal--Kubo formula holds true in general no matter whether the reference process with expectations $\langle \cdot\rangle_0$ is in equilibrium or  in some stationary nonequilibrium with density $\rho$.

As the simplest example we take a Langevin dynamics (and from now we put $k_B=1$) 
\begin{equation}\label{sq}
\dot{x}_s = \nu\, F(x_s) + \sqrt{2 \nu\, T}\,\xi_s
\end{equation}
for a single overdamped particle with position $x_t$ at time $t$ in a heat bath at temperature $T.$  In general  the mobility $\nu$ multiplying the force $F$ can also depend on the temperature. But that temperature dependence only gives rise to a  a mechanical-like perturbation which can be handled easily with ordinary path integral formalism. So, for the sake of simplicity throughout this paper we assume that the mobility $\nu$ (or damping $ \gamma$ in case of underdamped systems)   is temperature independent.

We also suppose that the force is sufficiently confining to establish a smooth stationary density $\rho$ satisfying the stationary Fokker-Planck equation $ L^{\dagger}\rho(x) =0$ (using a one--dimensional notation for simplicity), where
\bea
 L^{\dagger}\rho(x) \equiv -\frac{\partial}{\partial x}\{\nu\, F(x)\,\rho\}(x) + \nu\, T\, \frac{\partial^2}{\partial x^2}\rho(x) \n
\eea
The question of primary importance here is the response to a change in temperature $T\rightarrow T+\varepsilon.$  The Agarwal--Kubo formula \eqref{eq:ak} remains intact for such a thermal perturbation, i.e., nothing changes essentially with the perturbation in
\eqref{eq:pert} being
\begin{equation}\label{perth}
L^\dagger_\varepsilon \rho \equiv L^\dagger \rho + \varepsilon \nu\,\frac{\id^2 \rho}{\id x^2},\quad L_{\rm pert}^\dagger = \nu\,\frac{\id^2}{\id x^2}
\end{equation}
Thermal response is thus given through the Agarwal--Kubo formula in the seemingly simple expression
 \begin{equation}
\langle O(t)\rangle_\varepsilon - \langle O(0)\rangle_0 = \varepsilon \nu\,\int_0^t\, \left\langle \frac 1{\rho}\frac{\id^2 \rho}{\id x^2}(x_0)\, O(x_s)\right\rangle_0 \, \id s\label{eq:akthermal}
\end{equation}
which is absolutely well-defined and suffers no mathematical problems as long as $\rho$ is smooth and the process has integrable time-correlations.

Under detailed balance in \eqref{sq},  the force is derived from a potential, $F= -\id U/\id x$, and for reversible stationary, i.e., equilibrium density $\rho \sim e^{-\beta U}$, we have
(with $\beta= 1/T$, backward generator $L$ and $\langle\cdot\rangle_0 = \langle\cdot\rangle_{\text{eq}}$)
\begin{eqnarray}
\nu\frac 1{\rho}\frac{\id^2 \rho}{\id x^2} &=&-\nu\beta U''+ \nu(\beta U')^2 = -\beta^2 LU\label{equi}\\ 
Lf(x) &=& -\nu\frac{\id U}{\id x}\frac{\id f}{\id x}+ \nu T\, \frac{\id^2 f}{\id x^2}\nonumber\\ 
\langle Lf(0) g(s)\rangle_{\text{eq}} &=& \frac{\id}{\id s}\langle f(0)\,g(s)\rangle_{\text{eq}}\nonumber
\end{eqnarray}
Therefore, inserting \eqref{perth}--\eqref{equi} into \eqref{eq:akthermal} gives the equilibrium response for the energy,
\begin{equation}\label{eqres}
\langle U(t)\rangle_\varepsilon - \langle U \rangle_{\text{eq}} = \varepsilon\beta^2[ \langle U^2\rangle_{\text{eq}} - \langle U(0)U(t)\rangle_{\text{eq}}] = \frac{\varepsilon}{2}\langle S(t)^2\rangle_{\text{eq}}
\end{equation}
in terms of the entropy flux $S(t) \equiv \beta\,(U(0)-U(t))$.

 Clearly however, no such explicit computation works out of equilibrium except for special cases -- we do not know $\frac{\id^2 \rho}{\id x^2}/\rho$ in \eqref{eq:akthermal} or how to measure it, if we are truly away from equilibrium. In other words, we have no objections against the assumed smoothness but physically, the observable $L^\dag_{\rm pert} \rho / \rho$ featuring in the correlation functions \eqref{eq:ak} or \eqref{eq:akthermal} is not sufficiently explicit and is often of little practical use (however, formula \eqref{eq:ak} can be used for numerical approximations, for example via a fitting of $\rho$ \cite{Majda,Speck}). Moreover the Agarwal-Kubo scheme for perturbation is less adapted to observables like time-integrated currents that depend on the trajectory over multiple times; one needs a separate derivation of Green--Kubo relations.  Instead we prefer the set-up via dynamical ensembles that mathematically boils down to path-integration, that unifies Kubo with Green--Kubo relations and that 
does suggest a more 
powerful interpretation of the response formula; see e.g. the frenetic origin of negative differential response in \cite{bae13}.

\begin{figure*}[!t]
 \centering
\begin{tikzpicture}
\node (x) at (-1.6,0) {\large $T_L$};
\node (y) at (12.0,0) {\large $T_R$};
\node at (0,0.5) {$1$};
\node at (1.6,0.5) {$2$};
\node at (3.4,0.5) {$i$-$1$};
\node at (5.2,0.5) {$i$};
\node at (6.6,0.5) {$i$+$1$};
\node at (8.4,0.5) {$n$-$1$};
\node at (9.9,0.5) {$n$};
\draw[thick] (-1.1,-0.5) -- (-1.1,0.5);
\draw[thick] (11.4,-0.5) -- (11.4,0.5);
\fill [pattern = north east lines] (-1.3,-0.5) rectangle (-1.1,0.5);
\fill [pattern = north east lines] (11.4,-0.5) rectangle (11.6,0.5);
\node[circle,fill=blue,inner sep=1mm] (a) at (0,0) {};
\node[circle,fill=blue,inner sep=1mm] (b) at (1.6,0) {};
\node[circle,fill=blue,inner sep=1mm] (c) at (3.4,0) {};

\node[circle,fill=blue,inner sep=1mm] (d) at (5.2,0) {};
\node[circle,fill=blue,inner sep=1mm] (e) at (6.6,0) {};
\node[circle,fill=blue,inner sep=1mm] (f) at (8.4,0) {};
\node[circle,fill=blue,inner sep=1mm] (g) at (9.9,0) {};
\draw[thick,decoration={aspect=0.3, segment length=2mm, amplitude=2.5mm,coil},decorate] (-1.1,0) -- (a) ;
\draw[thick,decoration={aspect=0.3, segment length=2mm, amplitude=2.5mm,coil},decorate] (a) -- (b) ;
\draw[very thick,dotted](2.2,0) -- (3.0,0);
\draw[thick,decoration={aspect=0.3, segment length=2mm, amplitude=2.5mm,coil},decorate] (c) -- (d);
\draw[thick,decoration={aspect=0.3, segment length=2mm, amplitude=2.5mm,coil},decorate] (d) -- (e);
\draw[very thick,dotted](7.1,0) -- (7.9,0);
\draw[thick,decoration={aspect=0.3, segment length=2mm, amplitude=2.5mm,coil},decorate] (f) -- (g);
\draw[thick,decoration={aspect=0.3, segment length=2mm, amplitude=2.5mm,coil},decorate] (g) -- (11.4,0);
\end{tikzpicture}
\caption{Sketch of a chain of oscillators connected to two thermal reservoirs at temperatures $T_L$ and $T_R.$ 
\label{fig:chain}}
\end{figure*}
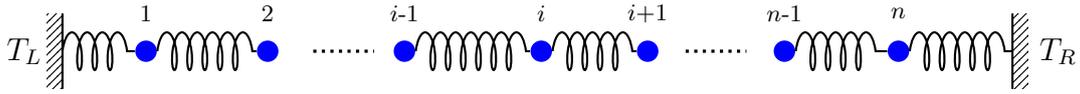

\subsection{The problem with path-integration}

The path-integration formulation allows for practically useful expressions for linear response formul{\ae}, readily applicable for nonequilibrium processes too~\cite{chat,bai09,bai09b,bai10}.
If one tries to apply that scheme to processes having different `temperatures,' problems of incommensurability arise.
In mathematics this is expressed by saying that the two processes are not absolutely continuous with respect to each 
other \cite{oks}. 
To illustrate the problem it suffices to inspect two oscillator processes for a single degree of freedom:
\[
\dot{x} = -\kappa_1 x + \sqrt{2D_1}\,\xi^{(1)}_s, \quad
\dot{y} = -\kappa_2 y + \sqrt{2D_2}\,\xi^{(2)}_s
\]
where $\xi^{(1)}_t$ and $\xi^{(2)}_t$ are two independent standard white noises.  If the diffusion constants $D_1=D_2$ are equal, then the two processes have the same support: their typical trajectories look the same and events that have zero probability for one have zero probability for the other process.  That is not true when $D_1\neq D_2$ for which sample paths lie in disjoint subsets of the set of all continuous trajectories.  An extreme example is $D_1=0$ and $D_2=D>0$ where the first motion would be exponentially decaying $x_t = x_0\exp -\kappa_1 t$, while the $y-$process clearly remains diffusive.  But even for $D_1>0$ and $D_1-D_2=\varepsilon\neq 0$ very small, the two motions remain mathematically mutually singular and there is no density of one with respect to the other process \cite{oks}.

To formally illustrate that problem in terms of path-integration, let us try to mimic the weight 
\[
\sim \exp\left[-\int_0^1 \frac{\dot{B}^2(s)}{4T}\id s\right]
\]
of a Brownian path $x_s = \sqrt{2T}B(s)$ at temperature $T$ on a discrete time grid.  Consider therefore a regular grid of mesh size $\Delta s =1/N$ in the unit time-interval $[t_0=0,t_N=1]$, and let us assign real variables $b_i$ to each time $t_i=0,1/N,2/N,\ldots,1$.  The Brownian weight resembles the (well-defined) density
\[
{\cal P}_T[b] = \left(\frac {N}{4\pi\, T} \right)^{N/2}\exp \left[-\frac{N}{4T} \sum_{i=0}^{N-1} (b_i-b_{i+1})^2 \right]\quad
\]
fixing $b_0=0$.
We recognize in the exponential a rescaled quadratic variation of a Brownian path $B(s)$.

Taking the derivative of the expected value for an observable $O(b)=O(b_1,b_2,\ldots,b_N)$ with respect to temperature we get the response formula
\begin{eqnarray}\label{ho}
&&\frac{\id}{\id T}\int_{R^N} \id b_1\id b_2 \ldots\id b_N\, O(b)\,{\cal P}_T[b] =\\
&& = \frac 1{2T}\int_{R^N} \id b\,\left[
\frac 1{2T}\sum_{i=0}^{N-1} \left(\frac{b_i-b_{i+1}}{1/N} \right)^2\frac 1{N} -  N \right] O(b)\,{\cal P}_T[b] \nonumber 
\end{eqnarray}
There, between $[\cdot]$, has appeared the rescaled quadratic variation  
\begin{eqnarray}
{\cal A}_N(b) &\equiv& \frac 1{2T}\sum_{i=0}^{N-1} \frac{(\Delta b_i)^2}{\Delta s} -  N  \\
&=& \frac 1{2T}\sum_{i=0}^{N-1} \left(\frac{b_i-b_{i+1}}{1/N} \right)^2\frac 1{N} -  N \nonumber
\end{eqnarray}
which has ${\cal P}_T$-mean zero, but its variance
\[
\int_{R^N} \id b_1\id b_2 \ldots\id b_N\, {\cal A}^2_N(b)\,{\cal P}_T[b] \propto N
\]
is diverging with $N\uparrow \infty$.  Clearly then, for some observables $O$ the response formula \eqref{ho} will stop making sense in the continuous time limit for $N\uparrow \infty$.  For other observables which are sufficiently localized or for which the quadratic variation converges to zero with $N$, we can hope there is a limit and that we can then exchange the $T-$derivative with the $N\uparrow \infty$ limit.  Simple examples of the latter are 'single-time' observables, like those $O$ considered in the previous subsection for the response \eqref{eq:akthermal}, or regular time-integrals of such observables.  For observables of the form
\[
O(b) = \sum_i f(b_i)  \,(b_{i+1} - b_i)
\]
which resemble stochastic integrals, the limit also works as long as the function $f$ is sufficiently smooth.

The above analogue inspires the remedy for our problem: first discretize and do the thermal response in a regularized version avoiding the singular behavior of white noise.  That is in fact what one is doing for discretization of the Langevin dynamics for numerical integration. 
For example, one can consider the Euler discretization scheme for a single underdamped particle with unit mass, in contact with a reservoir at temperature $T,$
\bea
\D x_s &=&  v_s \Delta s  \cr
\D v_s &=&   -  \gamma v_s\,\D s + \sigma \sqrt{\Delta s}~ \eta_s \label{eq:euler}
\eea
Here $\sigma=\sqrt{2\gamma T}$ and $\eta$ is a Gaussian random number with mean zero   and unit variance. The $\Delta$ refers to position, velocity and time increments; e.g.~$\Delta v_s = v_{s+ \Delta s} -  v_s$ for some very small $\Delta s>0.$  There are other, more accurate, discretization schemes too. To be specific we add another scheme \cite{Tuckerman,Ciccotti},   
\bea
 \label{eq:verlett}
\D x_s &=&  v_s \Delta s + \alpha(s) \\
\D v_s &=&  -  \gamma v_s\,\Delta s + \sigma\sqrt{\Delta s}~ \eta_s -\gamma \alpha(s) \cr
{\text{with}} &&\alpha(s) = - \gamma \frac {\Delta s^2}2\, v_s + \sigma \Delta s^{3/2} \left( \frac 12 \eta_s + \frac 1{2\sqrt 3}\theta_s  \right) \n
\eea 
Here $\sigma=\sqrt{2\gamma T}$ and $\eta$ and $\theta$ are independent Gaussian random numbers with $\la \eta \ra = \la \theta \ra=0$  and $\la \xi^2 \ra = \la \theta^2 \ra=1.$  It is easy to check that this converges to the traditional Langevin dynamics in the continuous time limit. 

It is possible to give the explicit path--weight $P(\D x_s,\D v_s)$ for a piece of trajectory in the discrete picture and to see how that changes under a temperature change $T\rightarrow T'$ at time zero.  That clearly is sufficient for writing the linear thermal response, as we will make more explicit in the following sections with the example of the above two discretization procedures.

\section{The result}\label{sec:result}

Chains of oscillators are a classical example of systems driven out of equilibrium by being in contact with several spatially well-separated heat baths at different temperatures~\cite{fpu,Dhar,Lepri}.
We use a model of this kind to illustrate the structure of our results, whose derivation follows in the next section.

\begin{figure*}[!t]
 \centering
 \includegraphics[width=8 cm]{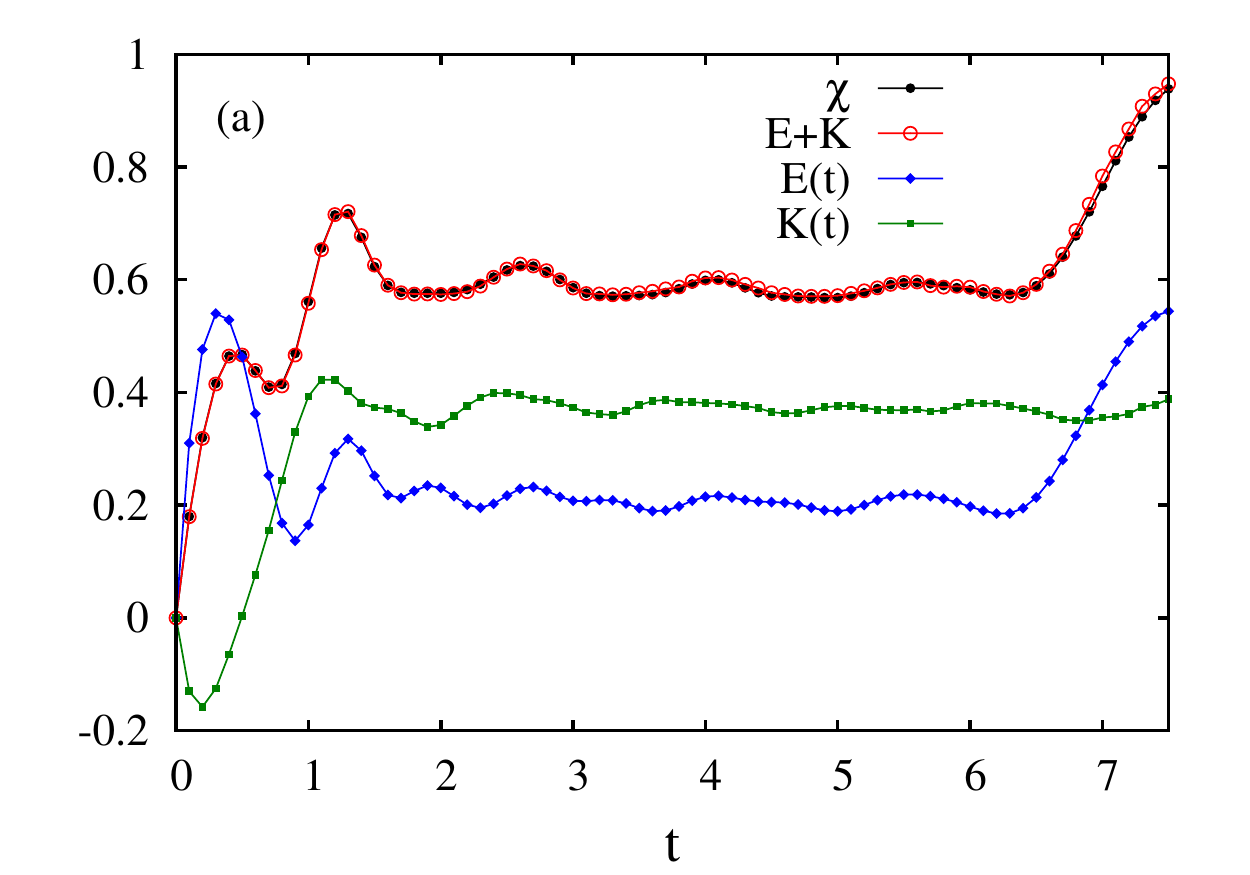} \includegraphics[width=7.1 cm]{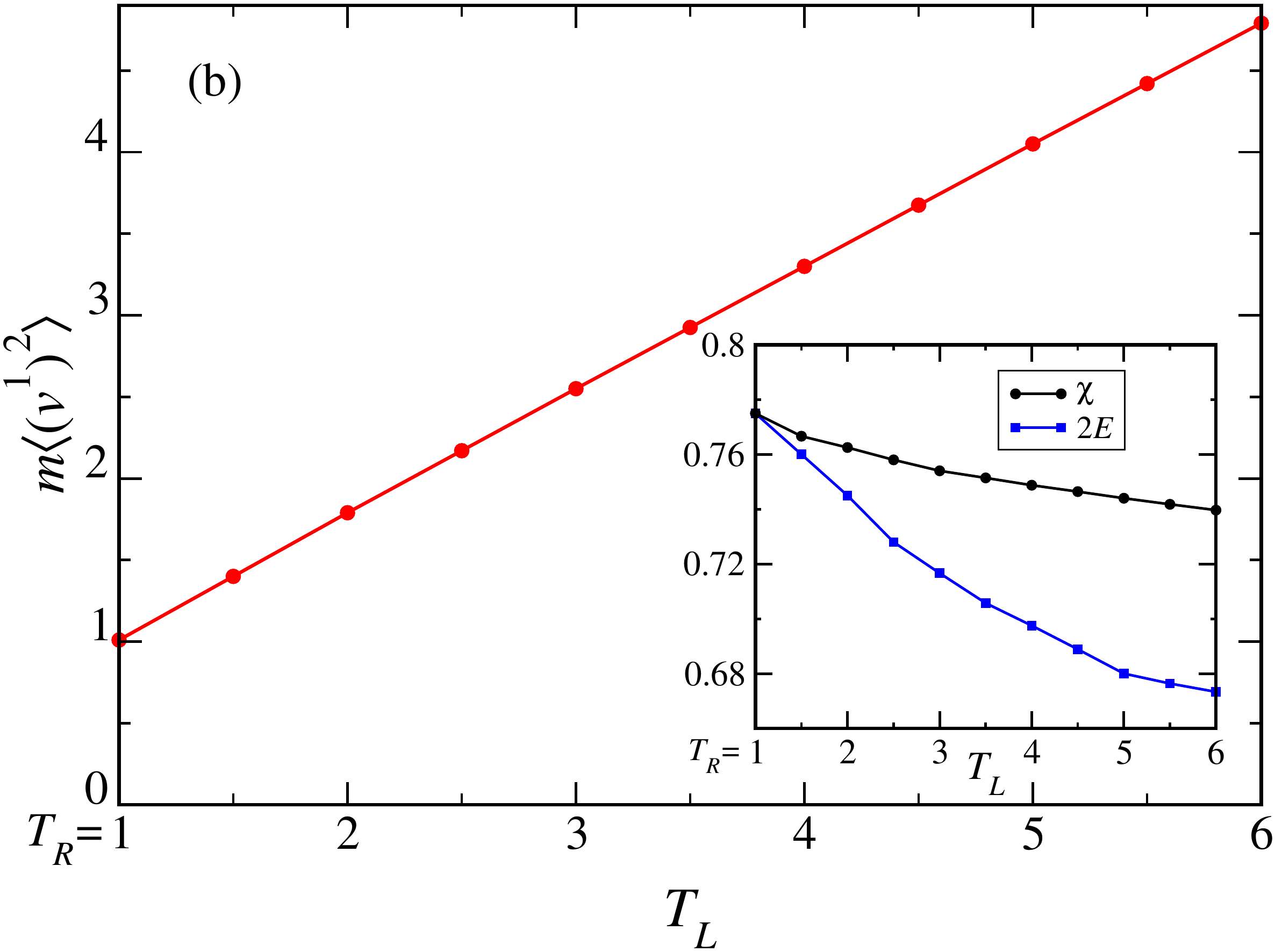}
 \caption{Thermal response of the kinetic temperature of the first oscillator  $m(v_t^1)^2$ in the open Fermi-Pasta-Ulam chain. (a) Plot of the susceptibility $\chi$ as a function of time for a chain of $n=11$ oscillators.  The red empty circles correspond to the response predicted by \eqref{eq:resp} and the black solid circles show the actual susceptibility measured numerically. The blue diamonds and green squares denote the entropic and frenetic contributions respectively. The right boundary reservoir has a fixed temperature $T_R=1.0.$ The left reservoir temperature is changed from  $T_L=2.0$ to $T_L' =2.2.$ (b)  The stationary values of the kinetic temperature of the first oscillator as the temperature of the left bath is changed from $T_L=2.0$ to $T_L=6.0$ keeping $T_R=1.0$ fixed. The inset shows the plot of the susceptibility $\chi$ (black  circles) and twice the entropic contribution $2E$ (blue squares). Here we have considered a chain of $n=7$ coupled oscillators. For both the plots (a) and (b) we 
have $\gamma_L=\gamma_R=1.0.$
  }\label{fig:FPU}
\end{figure*}

Take a chain of $n$ oscillators coupled to two thermal reservoirs with temperatures $T_L,T_R$ at the boundaries; see Fig.~\ref{fig:chain}. The position and velocity $(x^i,v^i)$  of the boundary oscillators evolve according to the underdamped Langevin equation,
\bea
\dot{x}_s^1&=& v_s^1, \quad m\dot{v}_s^1=F^1(x_s) - \gamma_L v_s^1 + \sqrt{2 \gamma_L T_L}~ \xi^{L}_s \cr
\dot{x}_s^n&=& v_s^n, \quad m\dot{v}_s^n=F^n(x_s) - \gamma_R v_s^n + \sqrt{2 \gamma_R T_R}~ \xi^{R}_s \label{eq:Lang}
\eea
while in the bulk there is a deterministic evolution
\bea
\dot{x}_s^i &=& v_s^i, \quad m\dot{v}_s^i=F^i(x_s) \quad \forall i \ne 1,n \n
\eea
The forces $F^i$ can contain both non-conservative and conservative parts. The noises $\xi_t^L, \xi_t^R$ are independent white noises and have the bath temperatures $T_L$ and $T_R$ in front of them.
We concentrate on fixing the friction coefficients $\gamma_L,\gamma_R$ and changing the temperature of the (say) left bath as $T_L \to T_L'$ at time zero where we start say from any arbitrary initial condition. 
Our result gives an expression for the thermal susceptibility of an observable $O,$ depending on the path $\omega$ (positions and velocities of all oscillators) in time-interval $[0,t]$
\bea
\chi_O \equiv \lim_{T_L'\to T_L}\frac{\la O \ra^{T_L'} - \la O\ra^{T_L} }{ T_L' - T_L} = E + K \label{eq:resp}.
\eea
$E$ and $K$ denote respectively the unperturbed correlations of the observable with excess entropy and dynamical activity:
\bea
E =-\frac 1{2 T_L}\left\langle O(\omega)\, ;\,S^L(\omega) \right\rangle^{T_L} \label{eq:entr}
\eea
where $\langle A ; B\rangle = \la AB \ra - \la A \ra \la B\ra$ is a connected correlation function, and $S^L(\omega)$ is the entropy flux into the left reservoir,
\[
S^L(\omega) = \frac 1{T_L} \left \{\frac 1{2}m(v_0^1)^2- \frac 1{2}m(v_t^1)^2 +\int_0^t \,F^1(x_s)\cdot v_s^1\,\id s \right\}
\]
The other term $K$ is time reversal symmetric and is termed the frenetic contribution.  The formal expression of $K$ depends on the discretization procedure used. Here we give an explicit form for the Euler scheme,
\bea
K  &=&  \frac{1}{4\gamma_L T^2_L}\,\int_0^t\id s \langle O(\omega)\, ;\,\{ (F^1)^2(x_s) - 2 m F^1(x_s) \dot{v}^1_s \}\rangle^{T_L} \nonumber \\ 
&& - \frac{\gamma_L}{2 mT_L^2} \int_0^t \id s \la O(\omega)\, ; \{T_L - \frac 12 m (v^1_s)^2 \} \ra^{T_L} \\
&& + \frac 1{2T_L^2}\lim_{\Delta s\downarrow 0} \left\langle O(\omega) \,;\,\sum_s \left\{-T_L + \frac{m^2}{2\gamma_L }\,\frac{(\Delta v_s^1)^2}{\Delta s} \right\} \right\rangle^{T_L} \nonumber \label{eq:loners}
\eea
where the sum $\sum$ is over the many time-steps in which $[0,t]$ is divided with mesh $\Delta s$.
That last term with $\sum\{-T_L + \frac{m^2}{2\gamma_L }\,\frac{(\Delta v^1)^2}{\Delta s}\}$
 is dangerously singular when split in two separate terms.  Yet, the combination $\frac{m^2}{2\gamma_L }\,\frac {(\id v^1_s)^2}{\id s} -T_L \sim \id s$ converges well in the time-continuum limit when evaluated in the correlation with physical observable $O$.

When the perturbation is around equilibrium, $T_L=T_R=T$ and all the forces are conservative, the  entropic and frenetic  contributions combine to make the Kubo formula
\bea
\chi^{eq}_O = 2E= -\frac 1T \la O(t)~ ;~ S^L(\omega) \ra \n
\eea
as follows in the usual way from symmetry arguments \cite{bai13}.

As an illustration we have measured the thermal  response of a boundary driven Fermi-Pasta-Ulam chain \cite{fpu,Dhar,Lepri} with interaction potential $V=\sum_{i=2}^n \frac 1{2} (x_i-x_{i-1})^2 + \frac 14 (x_i-x_{i-1})^4 ;$ the force acting on the $i^{th}$ oscillator is conservative in this case, $F^i(x) =- \frac{\partial}{\partial x_i} V(x)$, but a thermal difference $T_L \ne T_R$ keeps the system far from equilibrium. As an observable we choose  the kinetic temperature $O = m(v^1)^2$ of the leftmost oscillator. In Fig. \ref{fig:FPU}(a) we see the time-dependence of the response starting from an arbitrary state in which we fix $x_i=0,v_i=2\, \forall i$; both the susceptibility (red open circles) and the response predicted by  \eqref{eq:resp} (black filled circles) are measured. The entropic and frenetic components $E(t)$ (blue diamonds) and $K(t)$ (green squares) are also shown separately.  Fig. \ref{fig:FPU}(b) shows the asymptotic values ($t\uparrow \infty$) of the kinetic temperature as a function of the temperature of the left bath $T_L$ keeping $T_R$ fixed. We also plot in the inset  the susceptibility $\chi$ and twice the entropic  contribution $2E$ as a function of $T_L.$ 
The linear response regime around equilibrium, i.e., when $T_L=T_R$  we have $\chi= 2E,$   and the kinetic temperature almost equals the (left) temperature.  Further away from equilibrium, a heat current develops and the frenetic term  $K$ starts to play a bigger and separate role from the entropic contribution.

\section{The thermal response formula }\label{sec:thre}
Let us start by imagining a colloid of mass $m$ in a fluid at rest.  The colloid is undergoing an externally applied possibly  non-conservative force $F$. The work done is dissipated instantaneously as (Joule) heat to the fluid, which acts as a big thermostat, remaining by assumption in equilibrium at a fixed temperature $T$.  We can thus speak about its entropy and when the colloid at position $x_s$ moves with velocity $v_s$ at time $s\in [0,t]$, there is a time-integrated entropy flux
\bea
S = \frac 1{T} \left\{\frac 1{2}mv_0^2- \frac 1{2}mv_t^2 +\int_0^t \,F(x_s)\cdot v_s\,\id s \right \} \label{eq:S_underdamped}
\eea
(heat over temperature) spilled into the fluid.  That entropy flux plays a role in estimating the plausibility ${\c P}_T(\omega)$ of a path or trajectory
$\omega = (x_s, v_s, 0\leq s\leq t)$ with $\dot{x}_s=v_s$ started from a given initial condition $(x_0=x,v_0=v)$ for the colloid at time zero.  After all, from general principles of statistical mechanics summarized in the hypothesis of local detailed balance \cite{kls} we must have that 
\begin{equation}\label{ldb}
\frac{{\c P}_T(\omega)}{{\c P}_T(\theta \omega)} = e^{ S(\omega)}
\end{equation}
where $\theta \omega$ is the time-reversed trajectory.  We can thus write
\begin{equation}\label{nt}
{\c P}_T(\omega) = {\cal N}_T(\omega)\,  e^{ S(\omega)/2}
\end{equation}
where the prefactor ${\cal N}_T(\omega) = {\cal N}_T(\theta \omega)$ is time-symmetric, and expectations for a general path-observable $O$ of the colloid in $[0,t]$ are
\bea
\langle O \rangle_{x,v}^T &=& \int {\cal D}[\omega] \, {\c P}_T(\omega)\,O(\omega)\nonumber\\ 
\langle O \rangle^T &=& \int\,\id x\id v\,\mu(x,v)\,\langle O \rangle_{x,v}^T\nonumber
\eea
where ${\cal D}[\omega]$ is the formal volume element on path-space and $\mu$ is a probability density over the initial state possibly also depending on temperature.

 Slightly changing the temperature $T\rightarrow T'$ of the fluid for times $s>0$ and assuming that the fluid relaxes quasi--immediately to its new equilibrium, we will know the response of the colloid
 \beq
 \la O \ra^{T'} -\la O \ra^{T} \simeq
 (T'-T)  \int\,\id x\id v\,\mu(x,v)\,\frac{\id}{\id T}\langle O\rangle_{x,v}^T \label{eq:resp2}
 \eeq
  from the $T-$dependence in ${\c P}_T(\omega)$. The thermal response of $\la O\ra_{x,v}^T$ then follows from \eqref{nt},
 \begin{eqnarray}
\frac{\id}{\id T}\langle O(\omega)\rangle_{x,v}^T &=&  \frac 1{2}\left\langle O(\omega)\frac{\id}{\id T} S(\omega)\right\rangle^T_{x,v} \nonumber\\ &&+ \left\langle O(\omega)\,\frac{\id}{\id T}\log {\cal N}_T(\omega)\right\rangle_{x,v}^T \n
\end{eqnarray}
Taking $O=1$ in the above expression we get $\frac 12 \left \la \frac{\id}{\id T} S(\omega) \right \ra_{x,v}^T = \left \la \frac{\id}{\id T}\log {\cal N}_T(\omega) \right \ra_{x,v}^T.$ This allows for a more convenient expression involving connected correlations $\la ~ ; \ra$  (as in \eqref{eq:entr}),
 \begin{eqnarray}\label{fro1}
\frac{\id}{\id T}\langle O(\omega)\rangle_{x,v}^T &=&  \frac 1{2}\left\langle O(\omega)~;~\frac{\id}{\id T} S(\omega)\right\rangle^T_{x,v} \nonumber\\&&+ \left\langle O(\omega)\,~; ~\frac{\id}{\id T}\log {\cal N}_T(\omega)\right\rangle_{x,v}^T
\end{eqnarray} 
The question of thermal response is thus to understand the temperature dependence of $S$ and ${\cal N}_T$ in \eqref{nt}:
from \eqref{eq:S_underdamped}, the temperature dependence of the entropy is simply $\frac{\id}{\id T}S= -\frac 1T S.$ On the other hand, in general there will be  many kinetic details entering ${\cal N}_T$ making it largely intractable. Indeed, time-symmetric quantities like the collision frequency or mean free path will depend not only on the colloidal mass and size, on the forcing $F$ and on the density and the friction $\gamma$ in the fluid but also on its temperature. At this moment we can think of simple effective models like the Langevin evolution. 
For example, one can consider an underdamped motion,
\[
m\dot{v}_s = -\gamma\, v_s + F(x_s) + \sqrt{2D}\,\xi_s
\]
with $\xi_s$ being standard white noise responsible for the random force of the fluid on the colloid and we have joined $D=\gamma  T$ as an independent parameter.  It is then to be expected that
\bea
{\cal N}_T(\omega) = {\cal N}_T^0(\omega) \, \exp [-{\cal U}_F(\omega)] \label{eq:uF}
\eea
where ${\cal U}_F$ contains the effect of the force $F(x)$ on the time-reversal symmetric part of the path--probability. It is  calculable from the specific dynamics at hand (underdamped Langevin equation here) and does not pose any problem, as we will see in the next section. More ambiguities will arise from the term ${\cal N}_T^0$, which  is the expression of ${\cal N}_T(\omega)$ for $F=0$ (still depending on other parameters $\gamma$ and $D$). Using  \eqref{eq:uF} into \eqref{fro1} we get
\bea
\frac{\id}{\id T}\langle O(\omega)\rangle_{x,v}^T &=& 
\frac 1{2}\langle O(\omega)~; \frac{\id}{\id T} S(\omega)\rangle^T_{x,v} \nonumber\\&& 
- \la O(\omega)~;\frac{\id}{\id T}{\cal U}_F(\omega) \ra_{x,v}^T\nonumber\\&&
 + \la O(\omega)~; \frac{\id}{\id T}\log {\cal N}_T^0(\omega) \ra_{x,v}^T \label{key}
\eea
Hence, the regularization of thermal response is reduced to making sense of the last term, which is to find good path-integration approximations to the Ornstein-Uhlenbeck process $m\dot{v}_s = -\gamma\, v_s  + \sqrt{2D}\,\xi_s$ or, what amounts to the same, to make the appropriate discretization of Brownian motion (which corresponds to $\gamma=0, D > 0$) on path-space. 
Treating the motion in the overdamped limit meets similar problems, as shown next with an explicit calculation for a single overdamped particle.

\subsection{Overdamped motion}

The Langevin equation governing the position $x_t$ of an overdamped particle in a medium of uniform temperature $T$ is given by,
\bea
\dot{x}_s = \nu F(x_s) + \sqrt{2\nu T}~ \xi_s \n
\eea
$F(x_s)$ denotes the systematic force, be it conservative or non-conservative, acting upon the particle and the white noise $\xi_t$ signifies the random force. The constant $\nu$ is the mobility, assumed to be position and temperature independent for the sake of simplicity.

To explore the probability of a path $\omega = \{x_s; s \in [0,t]\}$ at a certain level of temporal coarse-graining we consider a discretized version of the Langevin equation where we split up the total time interval $t$ is split up into $N$ small but finite steps of duration $\D s$ with $t=N \D s.$  
The simplest possible discretization follows the so called `Euler scheme' where one writes, the increment in position during time step $\D s$
\bea
\Delta x_s =  \nu\, F(x_s) \Delta s + \sqrt{2 \nu\, T} \sqrt{\Delta s} ~ \eta_s \label{eq:euler}
\eea
Here $\eta$ is a Gaussian random variable with mean $0$ and unit variance. The probability for the increment $\Delta x_s$ can be found from the formal Gaussian weight of $\eta$
\beq
P(\Delta x_s) = \frac 1{\sqrt{4 \pi \nu T \Delta s }} \exp \left[- \frac{(\Delta x_s -\nu F(x_s) \Delta s )^2}{4  \nu T \Delta s}\right] \label{eq:over_Pt}
\eeq
The complete trajectory $\omega = \{ x_s\}$ over a time interval $[0,t]$ consists of $N$ such jumps; the continuum limit is the usual $\Delta s \downarrow 0, N \to \infty.$ The full path weight for this path $\omega$ can be considered as
\bea
{\c P}(\omega) = \prod_s P(\Delta x_s) \label{eq:full_path}.
\eea

In the spirit of the previous discussion, we rewrite the  probability of the full  path $\omega$ as,
\bea
{\c P}(\omega) = {\cal N}_T^0(\omega) \exp[S(\omega) /2] \exp \left[- {\cal U}_F \right] \label{eq:NT_S_U}
\eea
The entropy flux to the medium $S(\omega)$ along the path is given by the Stratonovich sum 
\bea
S(\omega) = \frac 1T \sum_s F(x_s) \circ \D x_s \n
\eea
over the discrete time steps. To extract the time-antisymmetric entropy part $S$ from \eqref{eq:over_Pt} we have used the conversion from It\^o to Stratonovich summing,
\bea
F(x_s) \circ \D x_s = F(x_s) \D x_s + \frac 12 \frac {\id F}{\id x} (\D x_s)^2 \n
\eea
to  leading order in $\D s.$ The force dependent part of the time-symmetric factor is then easily recognized,  
\bea
{\cal U}_F(\omega) = \frac{1}{4T}\sum_s\left\{\nu F^2(x_s) \D s + \frac{\id F}{\id x} (\D x_s)^2 \right \} \cr
\frac{\id}{\id T} {\cal U}_F(\omega) = -\frac{\nu}{4T^2}\sum_s \D s \left\{F^2(x_s)  + 2T \frac{\id F}{\id x}  \right \} \n
\eea
Note that we have used $(\D x_s)^2 \sim 2 \nu T \D s $  {\it  after } taking the derivative of ${\cal U}_F$ with respect to temperature. 

Both $S(\omega)$ and ${\cal U}_F(\omega)$ are well behaved functions and the limit $\D s \downarrow 0$ does not raise any problem.
That leaves the residual factor ${\cal N}_T^0(\omega),$ 
\beq
 {\cal N}_T^0(\omega) =\left(\frac 1{\sqrt{4 \pi \nu T \Delta s }}\right)^N \exp \left[-\frac 1{4\nu T} \sum_s \frac{(\D x_s)^2}{\D s} \right] \label{eq:NT_euler}
\eeq
where $N$ is the total number of discrete time steps that constitute the interval $[0,t].$  The important question remains  how to get a meaningful result from this apparently singular quantity in the limit  $\D s \downarrow 0.$ The answer is to first determine the response in the discrete picture and then take the continuum limit. From \eqref{eq:NT_euler},
\bea
\frac{\id}{\id T} \log {\cal N}_T^0(\omega) &=& 
\frac 1{2T^2} \left [-N T + \frac {1}{2\nu}\sum_s \frac{(\D x_s)^2}{\D s} \right] \nonumber\\
&=& \frac 1{2T^2} \sum_s  \left[ \frac {1}{2\nu}\frac{(\D x_s)^2}{\D s} -T  \right ] \n
\eea
Both the terms in the above expression are singular when considered separately but the combination $\frac {1}{2\nu}\frac{(\D x_s)^2}{\D s} -T \sim \D s $ as can be verified from \eqref{eq:euler}  and converges well in the $\D s \downarrow 0, N \to \infty$ limit.
Now we are allowed to take the time continuum limit and collecting all the pieces, we arrive at the final thermal response formula. In conclusion, the thermal susceptibility for the observable $O$ is given by  \eqref{eq:resp},
\bea
\chi_O \equiv \frac{\la O \ra^{T'} - \la O\ra^T }{ T' - T} = E + K \n
\eea
The term $E$ correlates the observable with the entropy in the unperturbed state,
\bea
E &=& - \frac 1{2T} \la O(\omega) ~; S(\omega) \ra^T \cr
&=& - \frac 1{2T^2} \left \la O(\omega) ~; \int_0^t F(x_s) \circ \id x_s \right \ra^T \n
\eea
The frenetic component is 
\bea
K &=&  \frac \nu {4T^2}\int_0^t \id s \left \la O(\omega)  ~;\left (F^2(x_s) 
 + 2 T \frac{\id F}{\id x} \right) \right \ra^T \nonumber\\ & &
+ \frac 1{2T^2}\left \la O(\omega) ~;  \lim_{\Delta s\downarrow 0} 
\sum_s \left (\frac {1}{2\nu}\frac{(\D x_s)^2}{\D s} -T  \right )\right\ra^T \n
\eea

One must remember that we have used a specific scheme \eqref{eq:euler} to discretize the Langevin equation. Even though the actual response would not depend on the discretization scheme, the formula might - that is to say the different terms in the action might have different expression depending on the particular discrete version used. This becomes more apparent in the next Section where we treat the thermal response of an underdamped particle with two different discretization schemes.

\subsection{Underdamped version}

The next step is to see how the analysis of the previous section generalizes to the underdamped situation. The particle of mass $m$ now has both a position and a momentum degree of freedom, with equation of motion 
\bea
\dot{x}_s = v_s, \; \quad \; m\dot{v}_s = F(x_s) - \gamma v_s + \sqrt{2 \gamma T}~ \xi_s \n
\eea
$\xi_t$ and $\gamma$ are the white noise and the friction associated with the thermal reservoir at temperature $T,$ respectively.  Trajectories $\omega = (x_s,v_s; 0 \le s \le t),$ are obtained in the discretized evolution with increments in position and velocity during time $s$ and $s+\D s$ given by
\bea
\D x_s &=& v_s \D s \cr
m\, \D v_s &=& F(x_s) \D s - \gamma v_s \D s +\sqrt{2 \gamma T} \sqrt{\D s}~ \eta_s \label{eq:euler_underd}
\eea
again using the Euler scheme. Since the position increment is completely determined by the velocity at the moment, the path weight for the piece of trajectory during time $s$ and $s+\D s$ satisfies $P(\D x_s, \D v_s) = P(\D v_s) \delta(\D x_s - v_s \D s).$ Then it suffices to inspect the path weight  $P(\D v_s).$   Following the exact same steps as the overdamped case, we identify the entropy generated along the full path $\omega,$ (taking already the limit $\Delta s\downarrow 0$)
\bea
S(\omega) = \frac 1T \left\{\int_0^t F(x_s) v_s \id s - \int_0^t v_s \circ \id v_s \right\} \n
\eea
as already written in \eqref{eq:S_underdamped}. The force dependence comes out to be 
\begin{equation}\label{split}
{\cal U}_F(\omega) = \frac 1{4 \gamma T}\int_0^t\id s\,\big(F^2(x_s) - 2m\,F(x_s)\dot{v}_s \big) 
\end{equation}
Once again the conversion from It\^o to Stratonovich 
\bea
v_s \circ \D v_s = v_s \D v_s + \frac 12 (\D v_s)^2 \n
\eea
has been used to identify the time-antisymmetric entropy flux.  While the entropy and the force-dependent part lend themselves directly to the continuum limit, one has to be careful regularizing the symmetric prefactor for $F=0,$
\bea
{\cal N}_T^0(\omega)&=& \left(\frac 1{\sqrt{4 \pi \gamma T \D s}}\right)^N 
\exp \left [-\frac 1{4 \gamma T} \sum_s \left\{m^2 \frac{(\D v_s)^2}{\D s } \right.  \right.  \n\\ 
&&\qquad \left.\left.+ \gamma^2 v^2 \D s - m\gamma (\D v_s)^2  \right\} \right] \n
\eea 

We calculate the change in this weight factor when the temperature is changed before taking the time continuum limit, and  the same structure as in the overdamped case can be recognized,
\bea
\frac{\id}{\id T}\log {\cal N}_T^0(\omega)& =& \frac 1{2T^2} \sum_s \left[\frac{m^2}{2 \gamma} \frac{(\D v_s)^2}{\D s} -T \right] \n\\
&& - \frac{1}{4T^2}\sum_s [m (\D v_s)^2  - \gamma v_s^2 \D s] \n
\eea
From the dynamics \eqref{eq:euler_underd}, $\frac{m^2}{2 \gamma} \frac{(\D v_s)^2}{\D s} -T \sim \D s$ and $ m^2 (\D v_s)^2 = 2 \gamma T \D s$ to first order in $\D s.$ Now we are allowed to take the limit $\D s \downarrow 0$ and piecing all the terms together in \eqref{key} and then using \eqref{eq:resp2}, the susceptibility for any observable $O$ is expressed as a sum of entropic and frenetic correlations as given by \eqref{eq:resp}. The entropic component is
\bea
E &=& -\frac 1{2T} \la O(\omega)~; S(\omega)\ra^T \n\\
&=& -\frac 1{2T^2} \left\la O(\omega) ~; \left\{\int_0^t F(x_s) v_s \id s - \int_0^t v_s \circ \id v_s \right\}\right \ra^T \n
\eea
and the frenetic component equals
\bea
K &=&  \frac{1}{4\gamma T^2}\,\int_0^t\id s \langle O(\omega)~; \{ F^2(x_s) - 2 m F(x_s) \dot{v}_s \}\rangle^T \n\\
&& - \frac{\gamma}{2 mT^2} \int_0^t \id s \la O(\omega)~; \{T - \frac 12 m v_s^2 \} \ra^T \cr
&& + \frac 1{2T^2} \lim_{\Delta s\downarrow 0}\left\langle O(\omega)~; \sum_s \left\{\frac{m^2}{2\gamma }\,\frac{(\Delta v_s)^2}{\Delta s} -T\right\}
 \right\rangle^T \n
\eea
where as usual correlations are measured in the unperturbed process.

To illustrate how the frenetic contribution depends on the discretization we take the other algorithm \cite{Ciccotti,Tuckerman} mentioned in the previous section,
\bea
\D x_s &=&  v_s \Delta s + \alpha(s) \\
\D v_s &=& \frac{\D s}2[F(x_s) + F(x_{s+\D s})] \cr
&&-  \gamma v_s\,\Delta s + \sigma\sqrt{\Delta s}~ \eta_s -\gamma \alpha(s) \cr
{\text{with}}\cr
\alpha(s) &=&  \frac {\Delta s^2}2\,(F(x_s) - \gamma v_s) + \sigma \Delta s^{3/2} \left( \frac 12 \eta_s + \frac 1{2\sqrt 3}\theta_s  \right) \n \label{eq:verlet_F}
\eea 
where we have assumed all masses $m=1$ for simplicity. The above dynamics emulates the same physical process described by the Langevin equation while offering the advantage  over the Euler algorithm of offering higher order corrections in $\D s$.  
The weight for a segment of path $(\D x_s, \D v_s)$ during time interval $\D s$ can be calculated from the probability distribution of the two independent Gaussian random numbers $\eta$ and $\theta.$ Casting the  weight of the full path into the form \eqref{eq:NT_S_U}, we have 
\bea
S(\omega) &=& \frac 1{2T} \sum_s [F(x_s) v_s \D s - \frac 32 v_s \circ \D v_s  + \frac{\D x_s \D v_s}{\D s}] \cr
&\simeq & \frac 1{2T}  \sum_s [F(x_s) v_s \D s - v_s \circ \D v_s ] \label{eq:sE_sv}
\eea
The last step follows from the dynamics \eqref{eq:verlet_F} to order $\D s.$ As expected, the expression for entropy remains same as in the Euler scheme. Also, ${\cal U}_F$ remains same as in \eqref{split}. The other factor ${\cal N}_T^0(\omega)$ however has a  different expression,
\bea
{\cal N}_T^0(\omega) &=& 
\left ( \frac{\sqrt 3}{2\pi \gamma T \D s^2}\right )^N \exp \left[-\frac 1T \sum_s \left \{\frac\gamma 4 v_s^2 \D s 
\right.\right.\cr &&\left. \left. 
+ \frac 3\gamma \frac{(\D x_s)^2}{\D s^3} - \frac 3{\D s^2} \left( (\D x_s)^2 + \frac 2\gamma v_s \circ \D x_s \right) 
\right.\right. \cr  && \left. \left.
+ \frac {6}{\D s} v_s \circ \D x_s  + \frac 1{4\gamma}\frac{(\D v_s)^2} {\D s}  +\frac 3{\gamma \D s} {\overline v}_s^2 - 3 {\overline v}_s^2  \right\} \right] \n
\eea
Here  ${\overline v}_s = v_s + \D v_s/2$ is the mean velocity during $\D s$, hence the Stratonovich product is discretized as $v_s \circ \D x_s \simeq {\overline v}_s\D x_s = v_s \D x_s + \frac 12 \D x_s \D v_s$.

The frenetic part of the linear  response formula \eqref{key} thus becomes 
\bea
K &=& \frac{1}{4\gamma T^2}\,\int_0^t\id s \langle O(\omega)~;\{ F^2(x_s) - 2 F(x_s) \dot{v}_s + \gamma^2 v_s^2 \}\rangle^T
 \cr &&
 +  \frac 1{T^2}  \left \la O(\omega)~; \lim_{\Delta s \downarrow 0}\sum_s \left\{\frac 3\gamma \frac{(\D x_s)^2}{\D s^3}
\right. \right. \cr && \left. \left.
 - \frac 3{\D s^2} \left( (\D x_s)^2 + \frac 2\gamma v_s \circ \D x_s \right) + \frac {6}{\D s} v_s \circ \D x_s 
\right. \right. \cr && \left. \left.
+ \frac 1{4\gamma}\frac{(\D v_s)^2} {\D s}  +\frac 3{\gamma \D s} {\overline v}_s^2 - 3 {\overline v}_s^2  -T \right\} \right \ra^T
\eea

In fact it contains a sequence of singular terms individually behaving like $\D s^0,$ $1/\D s$ and $1/\D s^2,$ which however combine  to result in a well behaved response. Moreover, as we said, for a given system the response has a unique value and it should not depend on the discretization scheme used to integrate the Langevin equation, hence the frenetic correlation $K$, even though very different formally,  must have the same value for same system parameters for all the discretization schemes, which we also checked numerically.  At any rate, the present solution in the treatment of thermal response for nonequilibrium systems, gives expressions like the ones above that appear to correspond to and are thus restricted to specific numerical schemes.  Obviously, when  the reference process is under equilibrium, the thermal response in the combination $E+K$ should again be given via the much more simple and universal \eqref{eqres}.
We have not investigated what the response formula becomes when the reference is close-to-equilibrium, and hence when the density in \eqref{eq:akthermal} can be approximated via a MacLennan--Zubarev form; see however \cite{kal} for such a study.

\subsection{Multiple temperature chains}

In general one is interested in systems composed by many degrees of freedom, 
some of which in direct contact with spatially separated heat reservoirs.
As long as all noise terms are statistically independent of each other,
one can simply add up contributions with the structure of the formul{\ae} presented for a single degree of freedom.
Of course, the contributions to consider are only those from the degrees of freedom in contact with the
altered reservoir.

\begin{figure}[t]
 \centering
  \includegraphics[width=8 cm,bb=0 0 360 252]{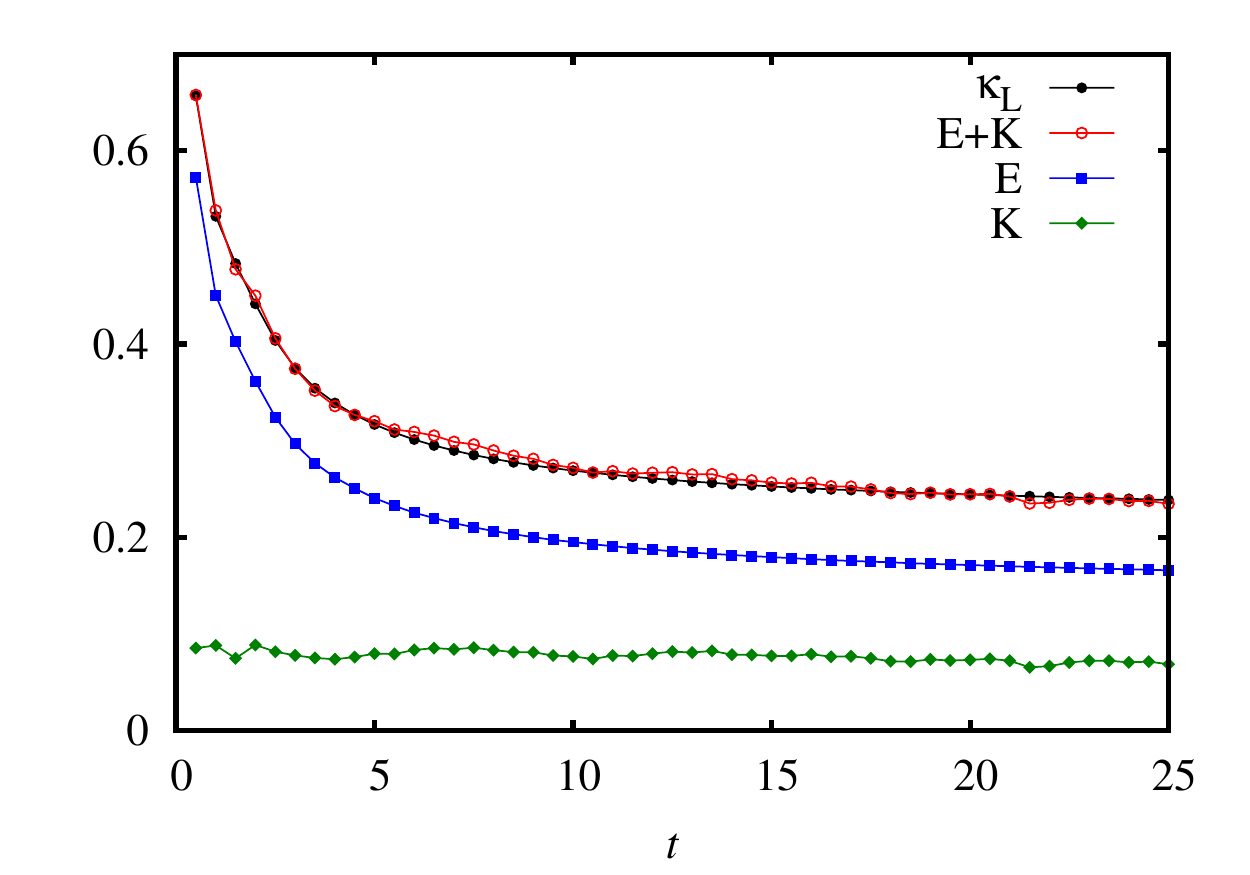}
 \caption{Response of stationary heat current of a chain of $n=11$ harmonic oscillators.
 The directly measured  susceptibility (black filled circles) matches very well with that obtained from the response formula (red empty circles). The entropic (blue squares)  and frenetic (dark green diamonds) components are also indicated separately.  Here  $T_R=1.0$ and $T_L=2.0$ is changed to $T_L'=2.2.$  Once again, $\gamma_L=\gamma_R =1.0$ are fixed.}
 \label{fig:heat_osc}
\end{figure}

As a general example we consider a chain of coupled oscillators with edges connected to two thermal reservoirs introduced in Section \ref{sec:result}. The goal is to predict the response of some observable when the temperature of one of the reservoirs is changed. Since the noise terms from the two baths are independent the path-weight can be expressed as products of the corresponding changes. The calculation  follows the same procedure as in the case of single particle, the only difference being that the relevant correlations are only with $v^1_t,$ the degree of freedom associated with the bath which is being perturbed, and we arrive at the result \eqref{eq:resp} - \eqref{eq:loners}.

In Section \ref{sec:result} we have given an example where the observable $O(t)$ only depends on the final time. An explicit path dependent observable is chosen here for further illustration. We look at the change in the average stationary heat current  flowing through the  left reservoir (which is same as the current flowing through the system in the stationary state) when the temperature of that reservoir is changed at time $t=0.$  In this case the observable is the  heat into the left reservoir per unit time  $O=j_h = T_L S^L/t.$ We choose a chain of harmonic oscillators; the system is described by the Langevin equations \eqref{eq:Lang} with $V=\sum_{i=2}^n \frac 1{2} (x_i-x_{i-1})^2$. The response of the heat current to a small change in the temperature of the left bath is the thermal conductivity $\kappa_L = \left. \frac{\partial j_h}{\partial T_L}\right|_{T_R}.$  Both the directly measured conductivity  (black dots) and that predicted  by the response formula (red empty circles) are shown in Fig. \ref{fig:heat_osc}. The corresponding entropic and frenetic components are  also  plotted in the same figure.

\section{Conclusions}
Thermal response for driven diffusive systems can be obtained from path integration methods under various time-discretization schemes.  There appears a rescaled quadratic variation of the process in a correlation function with the observation under consideration.  The time-continuum limit appears numerically stable when allowing enough sampling.  For the rest the thermal response follows the decomposition in an entropic and a frenetic contribution.  Not surprisingly, it is in the frenetic contribution that one finds the dangerously singular term reflecting the singular nature of white noise.
\\

\noindent {\bf Acknowledgments}: We thank Abhishek Dhar and Gianmaria Falasco for many helpful discussions. This work was financially supported by the Belgian Interuniversity Attraction Pole P07/18 (Dygest). We also thank the Galileo Galilei Institute for Theoretical Physics for the hospitality and the INFN for partial support during the completion of this work. Finally, M.B.~thanks ITF of KU Leuven for the hospitality and support.

\end{document}